\documentclass[aps,showpacs,twocolumn]{revtex4-2}
\usepackage[utf8]{inputenc}
\usepackage{amsfonts}
\usepackage{amssymb}
\usepackage{amsmath}
\usepackage{graphicx}
\usepackage{epsfig}
\usepackage{subfigure}
\usepackage{appendix}
\usepackage{hyperref}
\usepackage{color}

\hypersetup{hypertex=true, colorlinks=true, linkcolor=red, urlcolor=red, citecolor=red}

\begin{document}

\title{Magnetic Bloch oscillations in a non-Hermitian quantum Ising chain}
\author{K. L. Zhang}
\author{Z. Song}
\email{songtc@nankai.edu.cn}
\affiliation{School of Physics, Nankai University, Tianjin 300071, China}
\date{\today}

\begin{abstract}
We investigate the impacts of an imaginary transverse field on the dynamics of magnetic domain walls in a  quantum Ising chain. We show that an imaginary field plays a similar role as a real transverse field in forming a low-lying Wannier-Stark ladder. However, analytical and numerical calculations of the time evolutions in both systems show that the corresponding Bloch oscillations exhibit totally different patterns for the same initial states. These findings reveal the nontrivial effect of non-Hermiticity on quantum spin dynamics.
\end{abstract}

\maketitle

\section{Introduction}

\label{introduction}

Bloch oscillation (BO) describes the periodic motion of a wave packet
subjected to an external force in a lattice. This phenomenon was first noted 
by Bloch and Zener when they studied the electrical properties
of crystals \cite{bloch1929quantenmechanik, zener1934theory}. When an
external electric field is applied to a perfect crystal lattice, the
localized eigenstates with a ladder-like energy spectrum emerge, known
as the Wannier-Stark (WS) ladder \cite{wannier1960wave}. These states are closely
related to the BOs, which can be understood as the periodic motion of a wave
packet within the WS ladder, as an external field causes the
wave packet to transition between different WS states and exhibit
oscillatory behavior in terms of position and velocity. Experimentally, BOs were
observed in a semiconductor superlattice \cite{waschke1993coherent},
ultracold atoms in the optical lattice \cite{dahan1996bloch,
wilkinson1996observation, anderson1998macroscopic, morsch2001bloch} and many
other systems sequentially \cite{morandotti1999experimental,
sanchis2007acoustic, meinert2017bloch, zhang2022observation,
hansen2022magnetic}. It turns out that BO is a universal wave phenomenon. In 
the magnetic systems, BOs appear in the form of the magnetic domain-wall  
oscillations. As a nonequilibrium dynamic phenomenon in quantum many-body
systems, magnetic BOs in the quantum spin chains have attracted much 
attention from researchers \cite{kyriakidis1998bloch, sudzius1998optical, kosevich2001anomalous, cai2011quantum,
shinkevich2012spectral, kosevich2013magnon, shinkevich2013numerical,
syljuaasen2015dynamical, hansen2022magnetic}. Notably, inelastic neutron scattering experiments have provided evidence for the existence of magnetic BOs in the magnetically identical material $\mathrm{CoCl_{2}\cdot 2D_{2}O}$ \cite{hansen2022magnetic}.

In recent years, non-Hermitian physics have attracted much attention from various research areas \cite{longhi2009bloch, jin2010physics, lee2014heralded, jin2017topological, qin2020discrete, ashida2020non, jin2021symmetry, zhang2021observation, rohn2023classical, liang2023observation, xu2023pseudo, xu2023high}, and BOs have been investigated in a range of non-Hermitian
systems, including photonic lattices with gain or loss \cite{longhi2009bloch, longhi2009dynamic, parkavi2021stable}, tight-binding chains 
with an imaginary gauge field \cite{longhi2014exceptional, longhi2015bloch,
graefe2016quasiclassical}, and non-Hermitian frequency lattices induced by
complex photonic gauge fields \cite{qin2020discrete}. Classical systems such as 
photonics, mechanics and electrical circuits can be used to simulate
non-Hermitian wave physics at the single-particle level, while in the
quantum systems non-Hermitian Hamiltonians are mainly explained as the effective
descriptions of open quantum systems \cite{ashida2020non}, and have been experimentally 
realized in the systems of superconducting quantum circuits \cite%
{naghiloo2019quantum, partanen2019exceptional}, nitrogen-vacancy centers in
diamonds \cite{wu2019observation, zhang2021observation} and ultracold atoms \cite%
{li2019observation, ren2022chiral}. Moreover, it was proposed that an imaginary field in a spin chain can be implemented by a scheme similar to heralded entanglement protocols \cite{lee2014heralded}. More recently, researchers have shown that
complex fields in quantum spin models have unique impacts on the physical
properties of the systems \cite{zhang2020ising, liu2021quantum,
zhang2021quantum, lenke2021high, matsumoto2022embedding, guo2022emergent,
sun2022biorthogonal, gao2023experimental, lu2023unconventional, lenke2023series}, for
example, by driving a quantum phase transition and altering the phase diagram
of the system. 
However, to the best of our knowledge, the BO in a non-Hermitian quantum spin chain has not yet been
explored.

In this paper, we investigate the BOs of magnetic domain walls in a non-Hermitian quantum Ising chain.  
The model considered is a quantum Ising chain with real longitudinal and imaginary transverse fields. We show that in the small-field region, i. e., when the  strengths of two fields are much smaller than the Ising coupling, as well as in the $\mathcal{PT}$-symmetric parameters region that guarantees a full real spectrum, the low-energy dynamics of the magnetic domain walls are captured by a single-particle effective Hamiltonian, through which the physical mechanism of magnetic BOs is revealed. For real and imaginary transverse fields, the eigenstates of the effective Hamiltonian are both localized states with ladder-like energy spectra, forming the WS ladders. Analytical analysis and numerical calculation of the time evolutions show the occurrence of magnetic breathing and BO modes in the non-Hermitian quantum Ising chain by appropriately selecting the initial states.  It is shown that for the non-Hermitian quantum Ising chain, the dynamics for the Kronecker delta initial state is a breathing mode, while the Gaussian state remains stationary, which is totally different from the oscillation of the domain wall in a Hermitian quantum Ising chain. Interestingly, for the Bessel initial state, the BO mode appears, and the amplitude can be modulated by the strength of the  imaginary transverse field and the localization length of the initial state. 

This paper is organized as follows. In Sec. \ref{model}, we start by introducing the Hamiltonian of the quantum Ising chain with an imaginary transverse field, and derive the effective Hamiltonian and its solution. In Sec. \ref{magneticBOs}, we analyze the dynamics of BOs for three types of initial states on the basis of the effective Hamiltonian, while Sec. \ref{numerical_sim} presents the numerical results of the dynamics for the quantum spin chain. Finally, we conclude our findings in Sec. \ref{summary}.

\section{Model and effective Hamiltonian}

\label{model}

The model we consider is a quantum spin chain of length $N$ with the Hamiltonian 
\begin{equation}
H=H_{0}+H^{\prime },
\label{H_spin}
\end{equation}%
where
\begin{equation}
	H_{0} =-J\sum_{j=1}^{N-1}\sigma _{j}^{z}\sigma
_{j+1}^{z}-h_{z}\sum_{j=1}^{N}\sigma _{j}^{z}
\end{equation}
represents a spin chain in longitudinal magnetic field $-h_{z}$, and with ferromagnetic Ising coupling $-J$. For simplicity, we set $J=1$ in the following discussion. Here $\sigma _{j}^{\alpha }$ ($\alpha =x,y,z$) are the Pauli
operators on site $j$, while
\begin{equation}
	H^{\prime } =-g\sum_{j=1}^{N}\sigma _{j}^{x}
\end{equation}
is a transverse magnetic field term. In this paper, we consider both the Hermitian and non-Hermitian systems, when the transverse fields $g$ are taken as real and imaginary, respectively.  
We would like to point out that the Hamiltonian $H$ is different from that of the Yang-Lee Ising spin model, which exhibits  Lee-Yang zeros \cite{yang1952statistical, lee1952statistical, von1991critical, castro2009spin, ananikian2015imaginary, peng2015experimental, sanno2022engineering}; in this model, the longitudinal field is imaginary, and the transverse field is real instead. 
When $h_{z}=0$, the Hamiltonian $H$ reduces to the 
transverse field Ising chain, which is exactly solvable through the
Jordan-Wigner transformation when the periodic boundary condition is applied, 
and serves as a unique paradigm for understanding the quantum phase
transition \cite{pfeuty1970one}. A nonzero longitudinal field term involves nonlocal operators in the fermion representation, and thus breaks the solvability of the model.

\begin{figure}[t]
\centering
\includegraphics[width=0.5\textwidth]{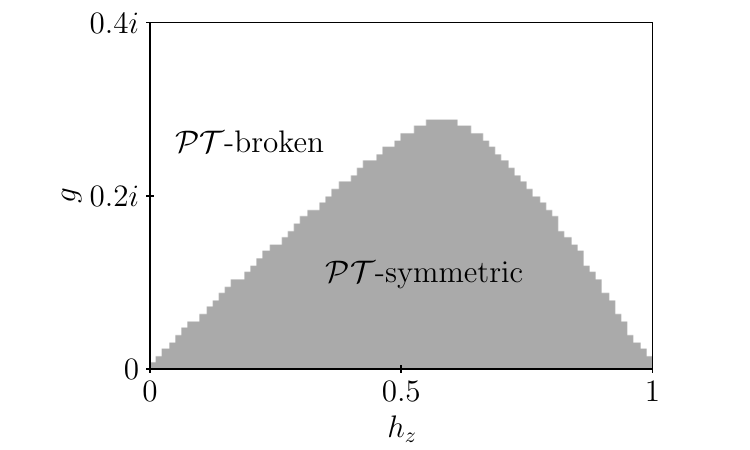}
\caption{Schematic of $\mathcal{PT}$-symmetric (gray) and $\mathcal{PT}$-broken (white) regions in the $h_{z}$-$g$ plane. The result is obtained by numerical exact diagonalization for a Hamiltonian $H$ with $J=1$ and $N=10$. }
\label{fig1}
\end{figure}

Imaginary fields in quantum spin systems have been discussed in many experimental and theoretical works \cite{lee2014heralded, wu2019observation, zhang2021observation, li2019observation, ren2022chiral, daley2014quantum, biella2021many, turkeshi2023entanglement}. In the framework of open quantum system dynamics, the imaginary transverse field $g$ arises in the no-click limit of the stochastic quantum jump trajectories when $( 1+\sigma^{x}_{j})  /2$ is measured \cite{daley2014quantum, biella2021many, turkeshi2023entanglement}, and $\left| g\right|  $ can be  interpreted as either dissipation or measurement rate. In this case, the resulted non-Hermitian Hamiltonian $H$ respects $\mathcal{PT}$ symmetry, that is, $\left[ \mathcal{PT},H\right]  =0$, with $\mathcal{P}=\prod^{N}_{j=1} \sigma^{z}_{j}$ being the parity operator, and $\mathcal{T}$ being the complex conjugation operator. This guarantees a full real spectrum in a certain region of system parameters \cite{bender1998real, bender1999PT, mostafazadeh2002pseudo, castro2009spin}, which is the so-called  $\mathcal{PT}$-symmetric region where the eigenstates remain unchanged under the action of the $\mathcal{PT}$ operator. Beyond this region, the eigenvalues occur in complex conjugate pairs and the corresponding eigenstates change under the action of the $\mathcal{PT}$ operator. Due to the lack of solvability in the model, the phase boundary of these two regions cannot be obtained analytically. Nevertheless, it is expected that the system possesses a full real spectrum when $\left| g\right|  $ is small compared to other system parameters. In Fig. \ref{fig1}, we presented the numerical result of the phase diagram of broken and unbroken $\mathcal{PT}$ symmetry for a finite-size system. The $\mathcal{PT}$-symmetric region is determined by condition $\sum^{2^{N}}_{n=1} \left| \mathrm{Im} \left( \varepsilon_{n} \right)  \right|  /2^{N}<0.01$, where $\varepsilon_{n}$ is the energy level of system $H$.

In this paper, we are concerned with the dynamics in the low-energy subspace as well as in the $\mathcal{PT}$-symmetric parameters region of the model. Thus, we content ourselves with a perturbation solution through seeking for an effective Hamiltonian describing the low-energy dynamics.
To proceed, we concentrate on the weak-field situation with $h_{z},\left\vert
g\right\vert \ll J$, and treat the transverse field term $H^{\prime }$ as a
perturbation in the following discussion.

We note that all the eigenstates of $H_{0}$ can be written in the tensor product form with fixed
numbers of spins that are parallel or antiparallel to the $z$ direction. The
ground state of $H_{0}$ is $\left\vert \Uparrow \right\rangle
=\prod_{j=1}^{N}\left\vert \uparrow \right\rangle _{_{j}}$ with energy $\mathcal{E}_{%
\mathrm{G}}=-N(J+h_{z})+J$. We focus on the low-energy subspace $\{\left\vert \phi _{m}^{\pm }\right\rangle \}$ that consists of states having one magnetic domain wall. Here $\left\vert \phi _{m}^{\pm }\right\rangle$ represent two types of domain-wall states:
\begin{equation}
\left\vert \phi _{m}^{+}\right\rangle =\prod_{l\leqslant m}\sigma
_{l}^{-}\left\vert \Uparrow \right\rangle ,\left\vert \phi
_{m}^{-}\right\rangle =\prod_{l>m}\sigma _{l}^{-}\left\vert \Uparrow
\right\rangle ,
\end{equation}
with $\sigma _{j}^{-}=(\sigma _{j}^{x}-i\sigma _{j}^{y})/2$ the lowering
operator, and $m=1,2,...,N-1$ the spatial position of the domain wall.
The corresponding energy is $\mathcal{E}_{m}^{\pm}=-N(J\pm h_{z})+3J\pm 2mh_{z}$. The action of$\ H$ on this basis yields 
\begin{eqnarray}
H\left\vert \phi _{m}^{\pm }\right\rangle &=&\left[ -N(J\pm h_{z})+3J\pm 2mh_{z}%
\right] \left\vert \phi _{m}^{\pm }\right\rangle \notag \\
&&-g\left\vert \phi _{m+1}^{\pm }\right\rangle -g\left\vert \phi _{m-1}^{\pm
}\right\rangle -g\left( ...\right) .  
\end{eqnarray}%
Here the ellipsis dots ``..." represent the terms containing the basis states with more than
one domain wall, which have at least $2J$ energy difference compared to the
states in $\{\left\vert \phi _{m}^{\pm }\right\rangle \}$. Thus, we are able to adiabatically 
eliminate these states and project the Hamiltonian $H$ into the subspace $%
\{\left\vert \phi _{m}^{\pm }\right\rangle \}$. The effective Hamiltonian is
given by \cite{sanz2016beyond} 
\begin{equation}
H_{\mathrm{eff}}=PHP-PHQ\frac{1}{QHQ}QHP,
\label{H_eff}
\end{equation}%
where the projectors are defined as $P=\sum_{m,\pm }\left\vert \phi _{m}^{\pm
}\right\rangle \left\langle \phi _{m}^{\pm }\right\vert $ and $Q=1-P$. The second term in Eq. (\ref{H_eff}) that is proportional to $g^{2}$ is discarded, considering the solvability of the effective Hamiltonian and $\left\vert g\right\vert $ is a small quantity. Up to first order, the
effective Hamiltonian has the explicit form
\begin{eqnarray}
&&H_{\mathrm{eff}} =-g\sum_{\lambda =\pm }\sum_{m=1}^{N-2}\left( \left\vert
\phi _{m}^{\lambda }\right\rangle \left\langle \phi _{m+1}^{\lambda
}\right\vert +\left\vert \phi _{m+1}^{\lambda }\right\rangle \left\langle
\phi _{m}^{\lambda }\right\vert \right)   \label{H_eff_1}\\
&&+\sum_{\lambda =\pm }\sum_{m=1}^{N-1}\left[ -N(J+\lambda h_{z})+3J+2\lambda mh_{z}\right]
\left\vert \phi _{m}^{\lambda }\right\rangle \left\langle \phi _{m}^{\lambda
}\right\vert . \notag
\end{eqnarray}

This indicates that the transverse field $-g$ acts as a hopping coefficient for the magnetic domain wall, while the strength of the longitudinal field $h_{z}$ plays the role of a skew potential.
Next, we investigate the dynamics in $\{\left\vert \phi _{m}^{+}\right\rangle \}$ subspace, and denote $\left\vert\phi _{m}\right\rangle =\left\vert \phi _{m}^{+}\right\rangle $  for simplicity. The analysis is similar for that of $\{\left\vert \phi _{m}^{-}\right\rangle \}$ subspace. 
In the absence of the skew potential $h_{z}$, the $k$-periodic spectrum is $E(k)=-2g\cos(k)+\textrm{const.}$ for the Bloch wave of magnetic excitation $\left\vert \phi \left( k\right)  \right\rangle  =$ $\left( 2\pi \right)^{-1/2}  \sum_{m} e^{imk}\left\vert \phi_{m} \right\rangle  $. For a real $g$, the semiclassical picture of BOs has been well understood \cite{bloch1929quantenmechanik, zener1934theory}. However, for an  imaginary $g$, the semiclassical picture should be understood in the framework of a modified equation of motion for expectation values, and the acceleration theorem holds only on average in time \cite{longhi2015bloch, graefe2016quasiclassical}.

The eigenstate of the Hamiltonian $H_{\mathrm{eff}}$ can be expanded as $\left\vert \psi
_{n}\right\rangle =\sum_{m}C_{m}^{(n)}\left\vert \phi
_{m}\right\rangle $, and the stationary Schr\"odinger equation $H_{\mathrm{eff}}\left\vert
\psi _{n}\right\rangle =E_{n}\left\vert \psi _{n}\right\rangle $ gives the
recursive relation for the expansion coefficients 
\begin{equation}
C_{m+1}^{(n)}+C_{m-1}^{(n)}=\frac{2\alpha _{m}}{z}C_{m}^{(n)},
\label{recursion}
\end{equation}%
with $\alpha _{m}=\left( 3J-NJ-E_{n}\right) /(2h_{z})+m-N/2$ and $z=g/h_{z}$%
. The boundary condition is $C_{0}^{(n)}=C_{N}^{(n)}=0$. We identify that
Eq. (\ref{recursion}) is the recursive formula of the Bessel function. Since the boundary effect is not involved in the dynamics that we will investigate in the next section, we assume an infinite chain in the following analytical analysis for convenience. Then the solution can be written as 
\begin{equation}
C_{m}^{(n)}=J_{m-n}\left( z\right),
\label{Cmn}
\end{equation}
which is the Bessel function of the first kind. Notably, the argument $z$ is imaginary for a non-Hermitian system. These eigenstates can be related by the spatial translation operation, that is, $\mathcal{T}\left\vert \psi_{n}\right\rangle=\left\vert \psi_{n+1}\right\rangle$ with the translation operator $\mathcal{T}$ defined as $\mathcal{T}\left\vert \phi_{m}\right\rangle=\left\vert \phi_{m+1}\right\rangle$.
 Then the eigenstates for the Hamiltonian $H_{\mathrm{eff}}$ are 
\begin{equation}
\left\vert \psi _{n}\right\rangle =\sum_{m}J_{m-n}\left(
z\right) \left\vert \phi _{m}\right\rangle ,
\label{eigenstates}
\end{equation}%
with energy $E_{n}=-N\left( J+h_{z}\right) +3J+2nh_{z}$, which is equally spaced and independent of $g$.

Similarly, the eigenstate of the Hamiltonian $H_{\mathrm{eff}}^{\dagger }$
with energy $E_{n}$ is
\begin{equation}
\left\vert \varphi _{n}\right\rangle =\sum_{m}J_{m-n}\left( z^{\ast }\right) \left\vert \phi
_{m}\right\rangle ,
\label{eigenstates_left}
\end{equation}
which establishes a biorthonormal basis set satisfying
\begin{eqnarray}
\langle \varphi _{n^{\prime }}\left\vert \psi _{n}\right\rangle &=&\delta
_{n^{\prime },n}, \\
\sum_{n}\left\vert \psi _{n}\right\rangle \left\langle \varphi
_{n}\right\vert &=&1.
\label{biorthonormal}
\end{eqnarray}

\begin{figure}[t]
\centering
\includegraphics[width=0.5\textwidth]{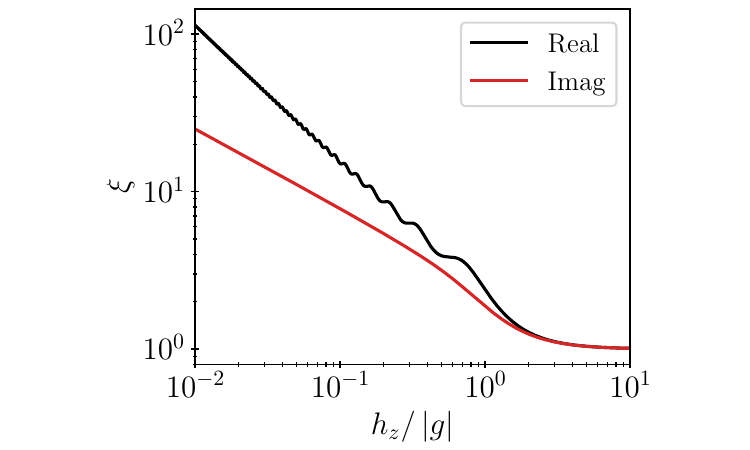}
\caption{Numerical results of the localization length $\xi$ defined in Eq. (\ref{localization_length}) as a function of $h_{z}$ (in units of $\left|g\right| $). The black and red lines represent the data for the real $g$ and imaginary $g$, respectively.  The horizontal and vertical axes are both on a logarithmic scale. The other parameters are set as $N=10^{4}$,  $J=1$, and $n=N/2$.}
\label{fig2}
\end{figure}

It is well known that the eigenstates are localized for a Hermitian WS ladder \cite{fukuyama1973tightly}. Thus, it can be reasonably inferred that this is also the case for an imaginary $g$. This can be confirmed  by the localization length for the eigenstate, which is defined as \cite{krimer2009delocalization, kosevich2013magnon}
\begin{equation}
	\xi =\frac{\left[ \sum_{m} \left| J_{m-n}(z)\right|^{2}  \right]^{2}  }{\sum_{m} \left| J_{m-n}(z)\right|^{4}  }  .
	\label{localization_length}
\end{equation}
For an infinite system, $\xi$ is independent of energy. Also, $\xi$ of the localized state is independent of $N$ when $N$ is large enough. In Fig. \ref{fig2}, we present the numerical results of the localization length $\xi$ of the eigenstates for the real and imaginary fields $g$, respectively. We can see that for both cases, a nonzero longitudinal $h_{z}$ field induces the localization of the eigenstates, which is more pronounced for an imaginary $g$. The localization of eigenstates is crucial for the upcoming discussion.

\section{Analyses for the oscillation dynamics}

\label{magneticBOs}

In this section, we investigate the dynamics of magnetic BOs in a non-Hermitian  quantum spin chain through an analytical analysis of the effective Hamiltonian. The characteristics of eigenstate localization and equal spacing of energy levels are both crucial for the construction of the initial excitation of the magnetic BOs.

We consider the initial states
\begin{equation}
	\left\vert \Psi (0)\right\rangle
=\sum_{m}f_{m}(0)\prod_{l\leqslant m}\sigma _{l}^{-}\left\vert \Uparrow \right\rangle
=\sum_{m}f_{m}(0)\left\vert \phi _{m}\right\rangle 
\end{equation}
 with three types of distributions representing the domain wall localized at site $m_{0}$: (i) the
delta function $f_{m}^{(1)}(0)$ $=\delta _{m,m_{0}}$; (ii) the broad
Gaussian distribution $f_{m}^{(2)}(0)$ $=\mathcal{N}^{-1}e^{-\alpha
^{2}(m-m_{0})^{2}+ik_{0}m}$ where $\mathcal{N}$ is a normalization coefficient, $%
\alpha $ characterizes the width of the distribution, and $k_{0}$ is the
wavevector; and (iii) the Bessel distribution $f_{m}^{(3)}(0)$ $=\Omega
^{-1}J_{m_{0}-m}\left( \kappa \right) $ with a complex argument $\kappa $.
According to the Schr\"odinger equation, the evolved state can be formally written as
\begin{eqnarray}
\left\vert \Psi (t)\right\rangle &=&\sum_{m}f_{m}(t)\left\vert \phi
_{m}\right\rangle \notag\\
&=&\sum_{m,m^{\prime }}K_{m,m^{\prime }}(t)f_{m^{\prime }}(0)\left\vert \phi
_{m}\right\rangle.
\label{Psi_t_f}
\end{eqnarray}

From the solution in Eqs. (\ref{eigenstates})-(\ref{biorthonormal}), the propagator $K_{m,m^{\prime }}(t)$ under
the biorthonormal basis can be computed as follows: 
\begin{eqnarray}
K_{m,m^{\prime }}(t) &=&\sum_{n}e^{-iE_{n}t}\left\langle m\right. \left\vert
\psi _{n}\right\rangle \left\langle \varphi _{n}\right\vert \left. m^{\prime
}\right\rangle \notag\\
&=&\sum_{n}e^{-iE_{n}t}J_{m-n}\left( z\right) J_{m^{\prime }-n}^{\ast
}\left( z^{\ast }\right),
\end{eqnarray}
and then using Graf's addition theorem \cite{abramowitz1968handbook, holthaus1996localization} for the Bessel functions in the summation
of index $n$, we
arrive at
\begin{equation}
K_{m,m^{\prime }}(t)=e^{i(\pi /2-h_{z}t)(m-m^{\prime })-i2m^{\prime
}h_{z}t}J_{m-m^{\prime }}\left[ \frac{2g\sin \left( h_{z}t\right) }{h_{z}}%
\right] .
\end{equation}
Here, a $m,m^{\prime }$-independent overall phase factor is discarded. Obviously, the propagator is periodic with a Bloch period $T=\pi /h_{z}$.

\subsection{Kronecker delta initial state}

Then, for the initial state $f_{m}^{(1)}(0)$ $=\delta _{m,m_{0}}$, the
evolved state is simply 
\begin{eqnarray}
f_{m}^{(1)}(t) &=&\sum_{m^{\prime }}K_{m,m^{\prime }}(t)\delta _{m^{\prime
},m_{0}}  \label{ft_1}\\
&=&e^{i(\pi /2-h_{z}t)(m-m_{0})-i2m_{0}h_{z}t}J_{m-m_{0}}\left[ \frac{2g\sin
\left( h_{z}t\right) }{h_{z}}\right] .\notag 
\end{eqnarray}
According to the properties of Bessel functions, the width of the domain wall periodically widens and narrows within the range 
\begin{equation}
\left\vert m-m_{0}\right\vert \lesssim \left\vert \frac{2g\sin \left(
h_{z}t\right) }{h_{z}}\right\vert ,
\end{equation}%
for both real and imaginary field $g$ with period $T=\pi /h_{z}$, which is the Bloch breathing mode. The profile of the evolved state  here is independent of the particular value of initial position $m_{0}$.

\begin{figure}[t]
\centering
\includegraphics[width=0.5\textwidth]{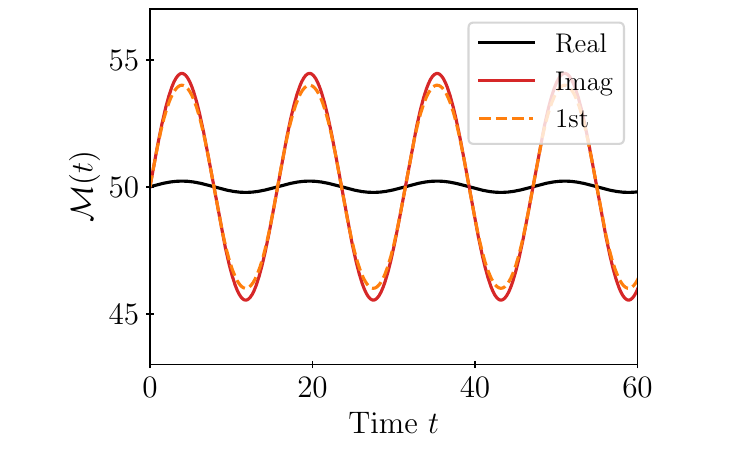}
\caption{Numerical results of the center of the wave packet as a function of time  defined in Eq. (\ref{COM}) for different fields: $g=0.1$ and $0.1i$ for the black and red lines, respectively. In addition, the orange dashed line represents the data obtained from Eq. (\ref{f3t}) with a first-order approximation for the case of $g=0.1i$. A comparison with the red line confirms the validity of the approximation. Other parameters are taken as $\kappa=10-0.5i$,  $N=100$, $m_{0}=50$, and $h_{z}=0.2$.}
\label{fig3}
\end{figure}

\begin{figure*}[t]
\centering
\includegraphics[width=0.9\textwidth]{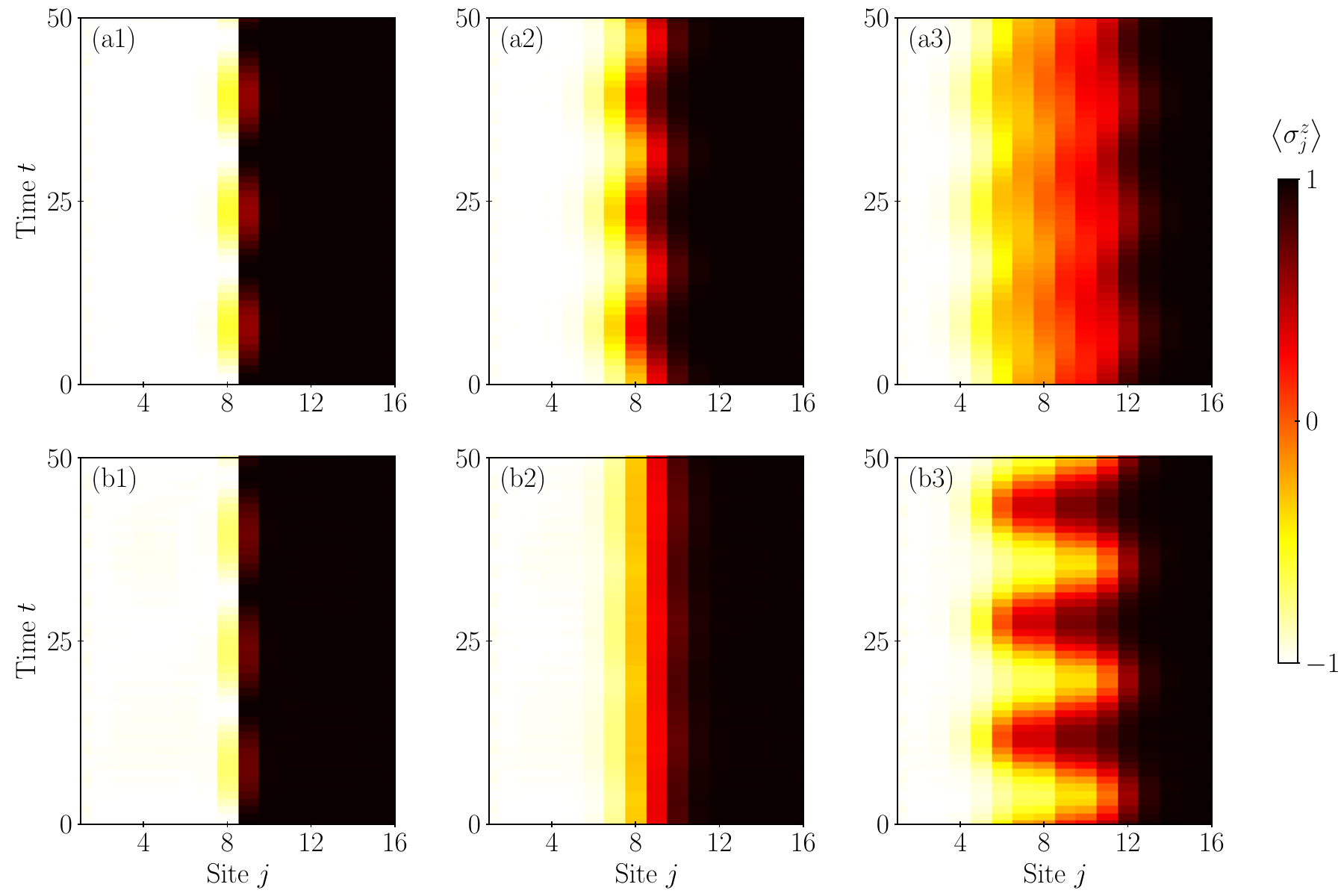}
\caption{Numerical simulations of the time evolutions for the three different initial states in the  
(a1)-(a3) Hermitian spin chain with $g=0.1$ and (b1)-(b3)  non-Hermitian spin chain with $g=0.1i$. In each figure, the expectation values of local spin $\sigma^{z}_{j}$ as functions of time are presented. The initial states in the left, middle, and right panels are taken as the Kronecker delta state,  Gaussian state with $\alpha=0.4$ and $k_{0}=0$, and Bessel state with $\kappa=4-0.5i$, respectively. Other parameters of the system: $N=16$, $J=1$, and $h_{z}=0.2$. }
\label{fig4}
\end{figure*}

\subsection{Gaussian initial state}

The evaluation of the time evolution for the initial state with Gaussian
distribution is not straightforward. Some approximations are needed. 
To do this, we first Fourier transform the time-evolution equation $%
f_{m}^{(2)}(t)=\sum_{m^{\prime }}K_{m,m^{\prime }}(t)f_{m^{\prime }}^{(2)}(0)
$ into $k$ space: 
\begin{eqnarray}
f_{k}^{(2)}(t) &=&\mathcal{N}^{-1}\sum_{m,m^{\prime }}e^{-ikm}K_{m,m^{\prime
}}(t)e^{-\alpha ^{2}(m^{\prime }-m_{0})^{2}+ik_{0}m^{\prime }} \notag\\
&=&\mathcal{N}^{-1}\exp \left[ \frac{2gi\sin \left( h_{z}t\right) \cos
(h_{z}t+k)}{h_{z}}\right] \notag\\
&&\times \sum_{m^{\prime }}e^{-\alpha ^{2}(m^{\prime
}-m_{0})^{2}-i(k-k_{0}+2h_{z}t)m^{\prime }}.
\end{eqnarray}
Assume that the spatial localization of the initial distribution is weak,
that is, $\alpha $ $\ll 1$, so that the summation of $m^{\prime }$ can be approximately 
replaced by integration. By doing this, we achieve
\begin{eqnarray}
&&f_{k}^{(2)}(t) \approx\frac{\sqrt{\pi }}{\alpha }\mathcal{N}^{-1}\exp %
\left[ \frac{2gi\sin \left( h_{z}t\right) \cos (h_{z}t+k)}{h_{z}}\right] \notag\\
&&\times e^{-i(k-k_{0}+2h_{z}t)m_{0}}\exp \left[ -\frac{(k-k_{0}+2h_{z}t)^{2}%
}{4\alpha ^{2}}\right] .
\end{eqnarray}
Again, since $\alpha $ $\ll 1$ is assumed, the momentum distribution $f_{k}(t)$ is
sharply localized around $k_{0}-2h_{z}t$. Then, we can expand the factor $\cos
(h_{z}t+k)$ in the argument of the exponential around $k_{0}-2h_{z}t$ up to the first order, and the
evolved state in real space can be obtained as
\begin{eqnarray}
&&f_{m}^{(2)}(t) =\frac{1}{2\pi }\int_{-\pi }^{\pi }e^{ikm}f_{k}(t)dk \label{ft_2}\\
&&\approx \mathcal{N}^{-1}\exp \left\{ i(k_{0}-2h_{z}t)m-i\Phi (t)-\alpha
^{2}\left[ m-M(t)\right] ^{2}\right\} ,\notag
\end{eqnarray}
where
\begin{eqnarray}
\Phi (t) &=&\frac{g}{h_{z}}\left[ \sin (k_{0}-2h_{z}t) - \sin k_{0}\right] , \\
M(t) &=&m_{0}+\frac{g}{h_{z}}\left[ \cos (k_{0}-2h_{z}t) - \cos k_{0}\right] .
\end{eqnarray}
For a real $g$,  the center of the wave packet $M(t)$ in real space oscillates in the form of a cosine function with period $\pi /h_{z}$ and amplitude $g/h_{z}$, which is the BO mode. However, for an imaginary $g$, the center of the wave packet remains stationary at the initial position $m_{0}$ for any initial wavevector $k_{0}$. Thus, in the following, we seek for a new initial excitation enabling magnetic BO to occur in the non-Hermitian quantum Ising chain.

\subsection{Bessel initial state}

Finally, we compute the time evolution for the initial Bessel distribution $%
f_{m}^{(3)}(0)$ $=\Omega ^{-1}J_{m_{0}-m}\left( \kappa \right) $ where $\kappa$ is a complex number characterizing the width of the initial distribution. Expanding this initial state  with the biorthonormal basis, the superposition coefficient is 
\begin{equation}
	 \langle \varphi_{n} |\Psi \left( 0\right)  \rangle =\Omega ^{-1}J_{m_{0}-n}\left( \kappa +g/h_{z}\right) . 
\end{equation}
For simplicity, we take $\kappa=x-g/h_{z}$ with a real $x$, then the above coefficient $\langle \varphi_{n} |\Psi \left( 0\right)  \rangle$ is always real.
The time evolution is computed as
\begin{eqnarray}
f_{m}^{(3)}(t) &=&\Omega ^{-1}\sum_{m^{\prime }}K_{m,m^{\prime
}}(t)J_{m_{0}-m^{\prime }}\left( x-g/h_{z} \right)  \label{bessel_f}\notag\\
&=&\Omega ^{-1}e^{-i2mh_{z}t}\left( \frac{x -ze^{-2ih_{z}t}}{x
-ze^{2ih_{z}t}}\right) ^{(m_{0}-m)/2} \notag\\
&&\times J_{m_{0}-m}\left( \sqrt{x ^{2}+z^{2}-2x z\cos (2h_{z}t)}%
\right) ,
\label{ft_3}
\end{eqnarray}
where $z=g/h_{z}$. 
Utilizing the multiplication theorem \cite{truesdell1950addition, abramowitz1968handbook} for the Bessel function, we obtain
\begin{eqnarray}
&&f_{m}^{(3)}(t) \notag\\&&=\Omega ^{-1}\sum_{n} \frac{1}{n!} \left[ \frac{ig\sin \left( 2h_{z}t\right)  }{h_{z}} \right]^{n}  J_{m_{0}-m+n}\left( x-ge^{-2ih_{z}t}/h_{z}\right) \notag\\ 
&& \approx  \Omega ^{-1}J_{m_{0}-m}\left( x-ge^{-2ih_{z}t}/h_{z}\right)  \notag\\
&&\quad+\Omega ^{-1}\frac{ig\sin \left( 2h_{z}t\right)  }{h_{z}} J_{m_{0}-m+1}\left( x-ge^{-2ih_{z}t}/h_{z}\right),
\label{f3t}
\end{eqnarray}
under the condition of $\left| g\right|  <h_{z}$.

While the results in Eqs. (\ref{bessel_f}) and (\ref{f3t}) indicate that this is a periodic oscillation with period $T=\pi/h_{z}$, the  pattern is not so explicit. Nevertheless, for a finite system size $N$, we introduce the center of the wave packet: 
\begin{equation}
\mathcal{M}(t)=	\frac{\sum_{m} m\left| f^{\left( 3\right)  }_{m}\left( t\right)  \right|^{2}  }{\sqrt{\sum_{m} \left| f^{\left( 3\right)  }_{m}\left( t\right)  \right|^{2}  } } .
\label{COM}
\end{equation}
The numerical results of $\mathcal{M}(t)$ for different values of $g$ are presented in Fig. \ref{fig3}. The figure shows that the center of the wave packet undergoes the BO over time, and the amplitude is on the order of magnitude $\left| \kappa \right|  $ for an imaginary $g$. However, for a real $g$, the center of the wave packet  remains near  the initial position $m_{0}$. This is opposite to that in the previous Gaussian initial state.

\section{Numerical simulations}
\label{numerical_sim}

Thus far, we have analyzed the time evolutions for three different initial excitations in the framework of the low-energy effective Hamiltonian in Eq (\ref{H_eff_1}).  It is worth noting that in the non-Hermitian system, the magnetic BO is absent for an initial Gaussian state but emerges for an initial Bessel state, which is distinct from the Hermitian system. In this section, we present the numerical simulations of the time evolutions for the three initial states under the original Hamiltonian in Eq. (\ref{H_spin}), in order to verify the previous analyses.

The initial states are taken as 
\begin{equation}
	\left\vert \Psi (0)\right\rangle=\sum_{m}f_{m}^{(n)}(0)\prod_{l\leqslant m}\sigma _{l}^{-}\left\vert \Uparrow \right\rangle
\end{equation}
with $n=1,\ 2$, and $3$, representing the Kronecker delta state, Gaussian state, and Bessel state, respectively, which are investigated in the previous section. The centers of these localized  initial states are all set as $m_{0}=8$, i.e., at the middle of the chain, to avoid boundary effects. 
 The evolved state $\left\vert \Psi (t)\right\rangle=e^{-iHt}\left\vert \Psi (0)\right\rangle/| e^{-iHt}\left\vert \Psi (0)\right\rangle |$ is calculated under the spin Hamiltonian in Eq. (\ref{H_spin}) using the fourth-order Runge-Kutta method with $5000$ time steps of length $\Delta t=0.01$, with a total accumulated error on the order of $O(\Delta t^4)$. The evolved state is normalized after each time step, and the local spin expectation value $\left< \sigma^{z}_{j} \right>  =\left< \Psi (t)\right\vert  \sigma^{z}_{j} \left\vert \Psi (t)\right\rangle  $ is computed after each 100 time steps. The results are presented in Fig. \ref{fig4}, and other parameters of the system and initial states are presented in the caption.

The boundary of  $\left< \sigma^{z}_{j} \right>=+1$ and $-1$ is the position of the magnetic domain wall. For the Hermitian spin chain, Figs. \ref{fig4}(a1)-(a3) show that the dynamics are magnetic Bloch breathing, BO, and stationary modes for the Kronecker delta, Gaussian, and Bessel initial states, respectively. 
With the same initial states, for the non-Hermitian spin chain, the results in Figs. \ref{fig4}(b1)-(b3) indicate that the dynamics are magnetic Bloch breathing, stationary, and BO modes, respectively. For the latter two initial states, the corresponding BOs exhibit totally different patterns for the same initial states in the two different systems. These numerical results are in accordance with the analyses in the previous section.

\begin{figure}[tbh]
\centering
\includegraphics[width=0.49\textwidth]{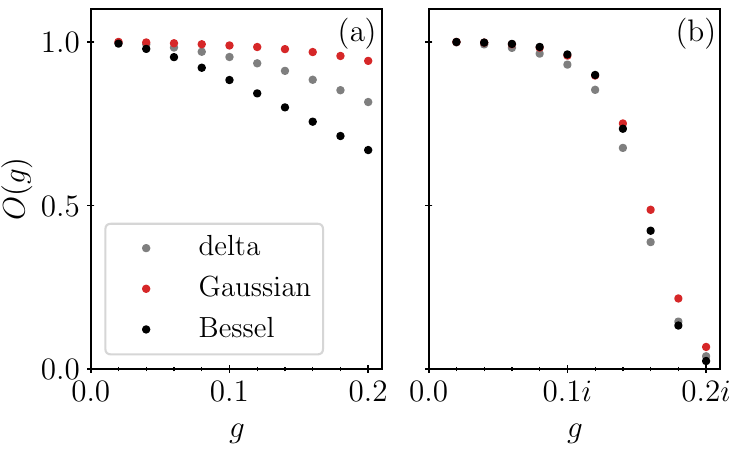}
\caption{Overlap of perturbative solutions and numerically evolved states at instant $t=10$ for three different initial states considered in the main text. (a) Overlap for the Hermitian system as a function of real transverse field $g$. (b) Overlap for the non-Hermitian system as a function of imaginary transverse field $g$. Parameters of initial states are taken as the same as those in Fig. \ref{fig4}. Other parameters of the system are set as $N=16$, $J=1$, and $h_{z}=0.2$.}
\label{fig5}
\end{figure}

In order to further estimate the validity of perturbative solutions presented in the previous section, we compare the numerical evolved states with perturbative solutions by the overlap
\begin{equation}
	O=\left| \langle \Psi_{\textrm{num.} } \left( t\right)  |\Psi_{\textrm{ana.} } \left( t\right)  \rangle \right|  
\end{equation}
at instant $t$, where $|\Psi_{\textrm{num.} } \left( t\right)  \rangle $ and $|\Psi_{\textrm{ana.} } \left( t\right)  \rangle $ denote the normalized numerical and analytical evolved states, respectively. The perturbative solutions $|\Psi_{\textrm{ana.} } \left( t\right)  \rangle $ for three initial states are taken as the form of Eq. (\ref{Psi_t_f}) with $f_{m}(t)$ being Eqs. (\ref{ft_1}), (\ref{ft_2}) and (\ref{ft_3}), respectively. In Fig. \ref{fig5}, we presented the numerical results of overlap $O$ at instant $t=10$ as a function of real and imaginary $g$. It indicates that for both cases, the perturbative solutions are in good agreement with the exact numerical results in a small $\left| g\right| $, while for imaginary $g$ the overlap drops sharply when $\left| g\right|  \gtrsim h_{z}/2 = 0.1$ due to the $\mathcal{PT}$ symmetry breaking of $H$ that is not captured by the effective Hamiltonian $H_{\textrm{eff}}$.

\section{Summary}
\label{summary}

In summary, we demonstrate the existence of the magnetic BOs in a non-Hermitian quantum Ising chain. It is shown that in the small-field region, the low-energy dynamics of the magnetic domain walls are captured by a single-particle effective Hamiltonian, with the transverse field acting as a hopping coefficient for the magnetic domain wall and the strength of the longitudinal field playing the role of a skew potential. For real and imaginary transverse fields, the eigenstates of the effective Hamiltonian are both localized states with equally spaced energy levels, forming the WS ladders. Analytical and numerical calculations of the time evolution for the non-Hermitian quantum Ising chain show the following.

(i) The dynamics of the Kronecker delta initial state follow a breathing mode.
 
(ii) The Gaussian state remains stationary, which is different from the oscillation of the domain wall in a Hermitian quantum Ising chain. 

(iii) For the Bessel initial state, the oscillation mode appears, and the amplitude can be modulated by the strength of the imaginary transverse field and the localization length of the initial state. 

The validity of perturbative solutions is estimated by comparing them with  numerical results. Our results reveal the mechanism of magnetic BOs in the non-Hermitian quantum spin chain and pave the way for future research on BOs in other quantum systems.

\acknowledgments This work was supported by the National Natural Science
Foundation of China (under Grant No. 12374461).

\end{document}